\documentstyle[epsfig,amssymb,prl,aps]{revtex}

\begin{document}
\title{Scaling of demixing curves and crossover from critical to tricritical
behavior in polymer solutions}
\author{J. S. Hager}
\address{Institute for Physical Science and Technology, \\
University of Maryland, College Park, Maryland 20742, U.S.A.}
\author{M. A. Anisimov, J. V. Sengers}
\address{Institute for Physical Science and Technology and Department of Chemical\\
Engineering, \\
University of Maryland, College Park, Maryland 20742, U.S.A.}
\author{E. E. Gorodetski\u{i}}
\address{Institute for Physical Science and Technology, \\
University of Maryland, College Park, Maryland 20742, U.S.A.\\
Oil and Gas Research Institute of the Russian Academy of Sciences,\\
Gubkina 3, 117971 Moscow, Russia}
\date{\today }
\maketitle

\begin{abstract}
In this paper we show that the virial expansion up to third order for the
osmotic pressure of a dilute polymer solution, including first-order
perturbative corrections to the virial coefficients, allows for a scaling
description of phase-separation data for polymer solutions in reduced
variables. This scaling description provides a method to estimate the $%
\Theta $-temperature, where demixing occurs in the limit of vanishing
polymer volume fraction $\phi $ and infinite chain-length $N$, without
explicit assumptions concerning the chain-length dependence of the critical
parameters $T_{{\rm c}}$ and $\phi _{{\rm c}}$. The scaling incorporates
three limiting regimes: the Ising limit asymptotically close to the critical
point of phase separation, the pure-solvent limit, and the tricritical limit
for the polymer-rich phase asymptotically close to the theta point. We
incorporate the effects of critical and tricritical fluctuations on the
coexistence curve scaling by using renormalization-group methods. We present
a detailed comparison with experimental and simulation data for
coexistence-curves and compare our estimates for the $\Theta $-temperatures
of several systems with those obtained from different extrapolation schemes.
\end{abstract}

%\begin{multicols}{2}

\section{Introduction}

\label{intro}

A quantitative description of the properties of a system in the vicinity of
multicritical points remains one of the most interesting problems in the
physics of phase transitions. Closeness to a multicritical point leads to a
complex crossover between several lines of critical points, which cannot be
described in terms of a single universality class. As a rule, the appearance
of a multicritical point on the phase transition line is related to the
interaction of two or more order parameters \cite{KS:84}. Correspondingly,
the complex crossover behavior is affected by a competition between the
diverging correlation lengths of the fluctuations of these different order
parameters. The most well-known example of such systems with a multicritical
(specifically, tricritical) point is the $^{3}$He-$^{4}$He mixture \cite
{KS:84,A:91}.

\begin{figure}[-t] \begin{center}
\epsfig{figure=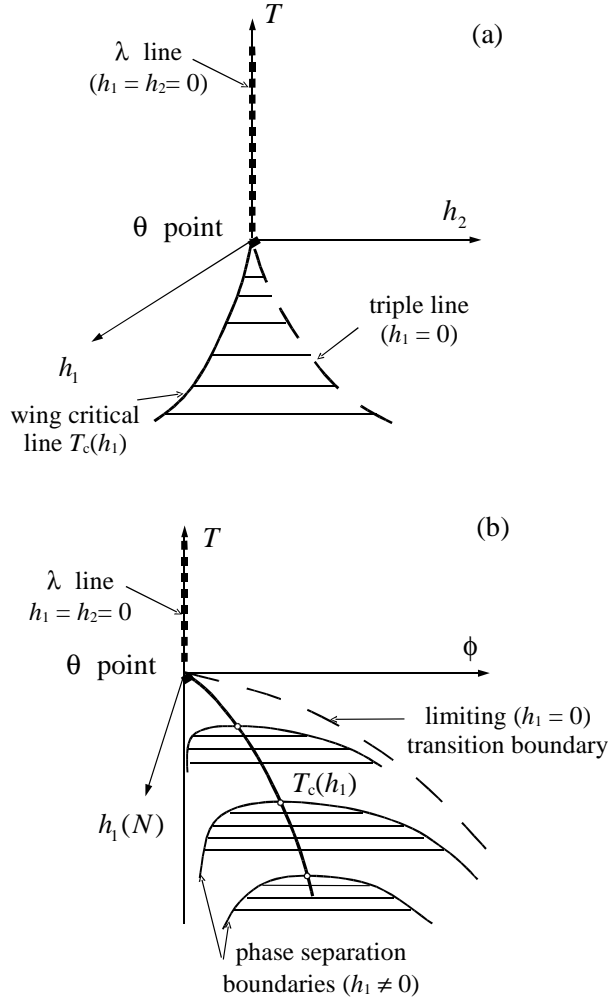,width=8cm} 
\vspace{0.5cm}
\caption[Tricritical phase diagram]{Schematic phase diagram near the
tricritical $\Theta $-point: (a) shown in field variables $T,h_{1},h_{2}$;
at $T=\Theta $ the $\protect\lambda $ line of self-avoiding-walk-type
critical behavior ends and a surface of first-order demixing transitions
with $h_{1}\neq 0$ emerges, which is bound by the wing line of critical
demixing points and by the triple line located at $h_{1}=0$; (b) replacing $%
h_{2}$ by the density variable $\protect\phi $ reveals two-phase
coexistence. Slices at constant chain length $N$, as indicated in the
figure, give the familiar phase-coexistence curves.}
\label{fig1}
\end{center} \end{figure}

In this paper we investigate the phase separation of solutions of
high-molecular-weight polymers in low-molecular-weight solvents in the
vicinity of the $\Theta$-point, which is defined as the critical mixing
point in the limit of infinite degree of polymerization \cite{deGennes}. De
Gennes \cite{dG:78}, by mapping Edwards' continuous chain model onto a
Euclidean field theory in the formal limit of zero spin components, has
argued that the $\Theta$-point in the polymer-solvent system is a
tricritical point. A tricritical point separates lines of second-order ($%
\lambda $-line) and first-order (triple-line) transitions. The states above
the $\Theta$-temperature on the line of zero polymer volume fraction $\phi
=0 $, shown by the dotted line in Fig. \ref{fig1}, correspond to the
critical self-avoiding-walk singularities associated with the behavior of
long $(N\rightarrow \infty )$ polymer molecules at infinite dilution \cite
{dG:75,F:94}. Later on, the mapping onto a field theory was generalized to
solutions of finite concentration and arbitrary polydispersity \cite
{deGennes,desC:75,M:77,SW:77,WP:81}. The scaling field ${\bf {h}_1}$,
conjugate to the polymer order parameter \boldmath$\psi\,$\unboldmath, is
zero along the $\lambda $-line but becomes non-zero for finite degrees of
polymerization $N$. The second scaling field $h_2$ (scalar) conjugate to $%
\psi^2$ also vanishes in the limit of infinite $N$ and zero polymer volume
fraction. The correlation length associated with the polymer order parameter
is proportional to the radius of gyration of a polymer which diverges in the
limit of infinite degree of polymerization. Below the (tricritical) $\Theta$%
-point the polymer order parameter exhibits a discontinuity accompanied by a
phase separation and by a discontinuity in the concentration of the polymer.
The line of critical phase-separation points shown in Fig. \ref{fig1} is a
nonzero-field critical (``wing'') line originating from the tricritical
point. The order parameter for the fluid-fluid phase separation, associated
with the volume fraction of monomers $\phi $, and the polymer order
parameter \boldmath$\psi\,$\unboldmath
belong to different universality classes. Tricriticality emerges as a result
of a competition between these two order parameters and exhibits mean-field
behavior with logarithmic corrections \cite{S:75,D:82,HS:99}.\ Specifying
the precise physical meaning of the polymer order parameter \boldmath$\psi\,$%
\unboldmath is a bit complicated (see Ref. \cite{deGennes} p. 287f). In
analogy to the $\lambda$-transition in $^4$He, one can view \boldmath$\psi\,$%
\unboldmath$({\bf {r})}$ as an operator for initiation or termination of a
polymer chain at a point ${\bf {r}}$, thereby relating it to the
concentration of polymer endpoints. On the level of two-point correlations,
one finds that the transverse correlations of \boldmath$\psi\,$\unboldmath
are related to the correlations of the ends of a single polymer, while the
longitudinal correlations (in the direction of ${\bf {h}_1}$) are related to
the correlations between all chain ends \cite{desC:75,M:77,SW:77}. A full
description of the phase separation near the tricritical point should
incorporate a crossover between Ising critical behavior on the wing critical
line and tricritical behavior close to the $\Theta$-point.

The separation of a polymer solution into two coexisting phases has been
qualitatively explained long ago by Flory and Huggins \cite{Fl:53}. However,
the Flory-Huggins theory describes the phase separation only qualitatively,
because it is based on a mean-field lattice model which neglects
fluctuations. It leads to the following Helmholtz free energy $F$ of mixing
of polymer chains and solvent molecules \cite{deGennes,DO:96}: 
\begin{equation}
\frac{F}{\Omega k_{{\rm B}}T}\equiv f=(1-\phi )\ln (1-\phi )+\frac{\phi }{N}%
\ln \phi +\frac{\Theta }{2T}\phi (1-\phi ),  \label{gflory}
\end{equation}
where $\Omega $ is the number of lattice sites in the system, $k_{{\rm B}}$
Boltzmann's constant, $T$ the temperature, while $\phi $ is the volume
fraction of the polymer, and $\Theta $ is the $\Theta $-temperature. We
obtain the Gibbs free energy $G$ via the Legendre transformation $G=F+PV$,
where $P$ is the pressure and $V=\Omega l^{3}$, with $l$ being the length of
a lattice cell. From the Gibbs free energy one can calculate the chemical
potential $\mu _{{\rm P}}$ of the polymer and $\mu _{{\rm S}}$ of the
solvent \cite{DO:96} 
\begin{eqnarray}
\hat{\mu}_{{\rm P}}(\phi ,P,T) &\equiv &\mu _{{\rm P}}(\phi ,P,T)-\mu _{{\rm %
P}}^{0}(T)=k_{{\rm B}}T\left( f+(1-\phi )\frac{\partial f}{\partial \phi }%
\right) +Pl^{3}\,,  \label{mupol} \\
\hat{\mu}_{{\rm S}}(\phi ,P,T) &\equiv &\mu _{{\rm S}}(\phi ,P,T)-\mu _{{\rm %
S}}^{0}(T)=k_{{\rm B}}T\left( f-\phi \frac{\partial f}{\partial \phi }%
\right) +Pl^{3}\,,  \label{musol}
\end{eqnarray}
which obey the relation $G/\Omega =\phi \hat{\mu}_{{\rm P}}+(1-\phi )\hat{\mu%
}_{{\rm S}}$. In Eqs. (\ref{mupol}) and (\ref{musol}) $\mu _{{\rm P}}^{0}(T)$
and $\mu _{{\rm S}}^{0}(T)$ are the standard chemical potentials of pure
polymer and solvent, respectively. The osmotic pressure $\Pi =-\hat{\mu}_{%
{\rm S}}/l^{3}+P$, due to the presence of the polymer, is the additional
pressure needed to establish equilibrium with the pure solvent across a
semipermeable membrane: 
\begin{equation}
\hat{\mu}_{{\rm S}}(0,P+\Pi ,T)=\hat{\mu}_{{\rm S}}(\phi ,P,T)\,.
\label{pi0def}
\end{equation}
Using Eq. (\ref{musol}) we find 
\begin{equation}
\hat{\Pi}\equiv \frac{\Pi l^{3}}{k_{{\rm B}}T}=\phi \frac{\partial f}{%
\partial \phi }-f\,.  \label{picalc}
\end{equation}
In the region of phase coexistence for $T<\Theta $, the values $\phi _{1}$
and $\phi _{2}$ of the volume fractions in the coexisting phases are found
from the condition of equal chemical potentials (or osmotic pressures) in
both phases: 
\begin{eqnarray}
\hat{\mu}_{{\rm S}}(\phi _{1},P,T) &=&\hat{\mu}_{{\rm S}}(\phi _{2},P,T)\,,
\label{cond1} \\
\hat{\mu}_{{\rm P}}(\phi _{1},P,T) &=&\hat{\mu}_{{\rm P}}(\phi _{2},P,T)\,,
\label{cond1b}
\end{eqnarray}
where the subscript 1 denotes the solvent-rich phase and 2 the polymer-rich
phase. Coexistence curves can also be calculated from the free energy by a
common-tangent construction 
\begin{equation}
\left. \frac{\partial f}{\partial \phi }\right| _{\phi _{1}}=\left. \frac{%
\partial f}{\partial \phi }\right| _{\phi _{2}}=\frac{f(\phi _{2})-f(\phi
_{1})}{\phi _{2}-\phi _{1}}\,,  \label{eqcond}
\end{equation}
which determines the two densities $\phi _{1}$ and $\phi _{2}$ of the
phase-separated system. At constant pressure and temperature these
conditions are equivalent to the equality of the chemical potentials of the
polymer and of the solvent in both phases. According to the Flory-Huggins
theory \cite{Fl:53}, the dependence of the critical temperature $T_{{\rm c}}$
and the critical volume fraction $\phi _{{\rm c}}$ of the polymer on the
degree of polymerization $N$ is $T_{{\rm c}}=\Theta /(1+1/\sqrt{N})^{2}$ and 
$\phi _{{\rm c}}=1/(1+\sqrt{N}),$ respectively. As shown by Widom \cite
{Wi:93}, for any value of the variable 
\begin{equation}
x=\frac{1}{2}\sqrt{N}(1-T/T_{{\rm c}})\,,
\end{equation}
but only if $N$ is large and $T$ is close to $T_{{\rm c}}$, the phase
coexistence in the Flory-Huggins theory can be represented by a scaling
form: the concentration difference $\phi _{2}-\phi _{1}$, where $\phi _{1}$
and $\phi _{2}$ are the volume fractions of polymer in the concentrated and
dilute phases, respectively, is given by 
\begin{equation}
\sqrt{N}(\phi _{2}-\phi _{1})\sim \left\{ 
\begin{array}{ll}
2\sqrt{6x} & (x\rightarrow 0) \\ 
3x & (x\rightarrow \infty )
\end{array}
\right. \,,
\end{equation}
where the limit $x\rightarrow 0$ corresponds to the critical point when $%
T\rightarrow T_{{\rm c}}$ for fixed $N$, and the other limit $x\rightarrow
\infty $ is approached when $N\rightarrow \infty $, for fixed $T$.

Asymptotically close to the critical point of phase-separation (the scaling
variable $x\rightarrow 0$) the Flory-Huggins theory fails, as does any other
mean-field theory which neglects critical fluctuations. The critical
behavior of the polymer solution near $T_{{\rm c}}$ is known to belong to
the three-dimensional Ising universality class. In particular, in the
asymptotic vicinity of the phase-separation critical point the concentration
difference $\phi _{2}-\phi _{1}$ as a function of the reduced temperature
difference $\tau =(T-T_{{\rm c}})/T_{{\rm c}}$ obeys a power law: 
\begin{equation}
\phi _{2}-\phi _{1}\simeq 2B_{0}\left| \tau \right| ^{\beta }\,.
\label{ising}
\end{equation}
However, Ising-like critical behavior is revealed only when the correlation
length of the critical density fluctuations is much larger than the radius
of gyration of the polymer chain \cite{MH:97,KA:02}. This fact implies that
the Ising region in high-molecular-weight polymer solutions is confined to a
very narrow temperature range near the critical point. Outside this region,
a crossover to mean-field type behavior is observed. A Flory theory
renormalized by critical fluctuations has been developed by Povodyrev {\it %
et al.} \cite{PS:99}. The extent of the Ising region is predicted to narrow
with increasing $N$ in a manner governed by the Ginzburg criterion, and
disappears entirely in the limit of infinite $N$. For large $N$, the
interplay between $\left| \tau \right| $ and $1/\sqrt{N}$ will drive the
system from asymptotic Ising critical behavior, when $|\tau |\sqrt{N}\gg 1$,
to asymptotic tricritical behavior, $\left| \tau \right| \sqrt{N}\ll 1$,
through a region of intermediate crossover behavior.

The approach developed in Ref. \cite{PS:99} contains most essential features
of real systems, namely: the crossover from mean-field to asymptotic
Ising-like behavior, the crossover to the (tricritical) $\Theta $-behavior,
and the variation of the effective critical exponents upon changing the
degree of polymerization. However, the restrictions imposed by the free
energy (\ref{gflory}) of the Flory-Huggins theory are too tight to represent
actual data within experimental accuracy. This is why here we shall use a
more general approach based on the virial expansion of the osmotic pressure $%
\Pi$, which we shall renormalize so as to include the effects of
fluctuations. We then try to describe existing experimental and simulation
data on phase separation in polymer solutions with this approach.

This paper is organized as follows. In Sec. \ref{meanfield} we briefly
review the connection between polymer solutions and field theory and show
that the Helmholtz free energy has a scaling form in the mean-field
approximation and also if one includes contributions from the first-order
perturbation theory. In Sec. \ref{critren} we describe how critical
fluctuations renormalize the free energy and in Sec. \ref{trisec} we
incorporate the effects of tricritical fluctuations. Sec. \ref{comp}
contains a detailed comparison of the renormalized theory with the
phase-separation data available from experiments and simulations. In Sec. 
\ref{concl} we summarize our findings.

\section{Mean-field description}

\label{meanfield}

Utilitising des Cloizeaux's mapping between Edwards' model of polymer
solutions in the case of nonzero concentration $\phi $ and a $O(n)$
symmetric field theory in a magnetic field of magnitude $h_{1}$ in the limit 
$n\rightarrow 0$, we find the following relations between the polymer
observables $\phi ,\hat{\Pi},c_{{\rm p}}=\phi /N$ and the effective
potential $\Gamma (\psi ,h_{2})$ of the field theory \cite
{deGennes,desC:75,M:77} 
\begin{eqnarray}
\phi &=&\frac{\partial \Gamma }{\partial h_{2}}\,,  \label{phidef2} \\
c_{{\rm p}} &=&\frac{\psi }{2}\frac{\partial \Gamma }{\partial \psi }\,,
\label{cpdef} \\
\hat{\Pi} &=&\psi h_{1}-\Gamma \,,  \label{pidef}
\end{eqnarray}
where $c_{{\rm p}}$ is the concentration of polymer molecules, $N$ is the
mean degree of polymerization. For the effective potential $\Gamma $, we
start from the mean-field expression 
\begin{equation}
\Gamma =h_{2}\psi ^{2}+\lambda \psi ^{4}+v\psi ^{6}\,,  \label{gammamf}
\end{equation}
which one may interpret as a Landau expansion in terms of $\psi $, but which
also emerges as the result of a zero-loop (``tree'') approximation in a
field theory or in the Edwards model \cite{D:82}. We should mention, that
des Cloizeaux's mapping \cite{desC:75} demands a polydisperse grand
canonical ensemble with an exponential chain-length distribution, while the
experiments that we want to describe are performed with nearly monodisperse
samples. A broad chain-length distribution in the experiments would
seriously affect the demixing transition, since the chain-length
distributions may differ in the phases below the demixing temperature, and
thereby may even change the nature of the transition. Polydispersity effects
have been incorporated in the theoretical framework for the excluded-volume
region, where the effective monomer-monomer interactions are repulsive and
therefore $\lambda $ is positive \cite{SW:80,lbook}. Fortunately, in the
tree approximation the effective potential $\Gamma $ and thereby the osmotic
pressure $\hat{\Pi}$ are independent of polydispersity. Also the first-order
correction for $\hat{\Pi}$ yields a change of only about $12\%$ when
changing from a monodisperse to an exponential chain-length distribution 
\cite{D:82}. Therefore, using the mean-field approximation for $\Gamma $, we
confidently apply the theory to experiments on nearly monodisperse solutions.

Minimizing Eq. (\ref{pidef}) for the osmotic pressure $\hat{\Pi}$ with
respect to $\psi $ we find 
\begin{equation}
h_{1}=\frac{\partial \Gamma }{\partial \psi }=\psi (2h_{2}+4\lambda \psi
^{2}+6v\psi ^{4})\,,  \label{h1vonphi}
\end{equation}
for the field $h_{1}$ conjugate to the order parameter $\psi $. Evaluation
of (\ref{phidef2}) and (\ref{cpdef}) gives 
\begin{eqnarray}
\phi &=&\psi ^{2}  \label{phivonpsi} \\
h_{2} &=&\frac{1}{N}+2\lambda \psi ^{2}+3v\psi ^{4}\,.  \label{h2vonphi}
\end{eqnarray}
Using (\ref{phivonpsi}) and (\ref{h2vonphi}) in Eq. (\ref{pidef}) we arrive
at the familiar expansion of the osmotic pressure near the $\Theta $-point 
\begin{equation}
\hat{\Pi}=\frac{\phi }{N}+\lambda \phi ^{2}+2v\phi ^{3}\,.  \label{exp_phi}
\end{equation}
Equation (\ref{exp_phi}) is the virial expansion of the equation of state.
Note that integration of Eq. (\ref{picalc}) using Eq. (\ref{exp_phi}) for $%
\hat{\Pi}$ leads to a free energy 
\begin{equation}
f=\frac{\phi }{N}\ln \phi +\lambda \phi ^{2}+v\phi ^{3}\,,  \label{helmexp}
\end{equation}
similar to Flory's expression (\ref{gflory}), with the $(1-\phi )\ln (1-\phi
)$ term expanded and with the expansion constants replaced by two
system-dependent interaction parameters $\lambda $ and $v$. The parameters $%
\lambda $ and $v$ are expected to be analytic functions of the temperature
with $\lambda $ changing its sign at the $\Theta $-temperature, leading to
the polymer-chain-collapse transition, while in the tricritical scenario $v$
is positive in order to stabilize the system at a finite density. To lowest
order we can write 
\begin{equation}
\lambda =\lambda _{0}\frac{T-\Theta }{\Theta }\,,  \label{lvont}
\end{equation}
where $\lambda _{0}>0$ is a system-dependent constant, so that $\lambda $ is
negative below the $\Theta $-point and is positive above the $\Theta $%
-point. The lowest-order approximation for $v$ is simply a constant
independent of temperature.

The lambda line is reached for $T>\Theta $ and $h_{1}=h_{2}=0$ (see Fig. \ref
{fig1}). At the triple line (coexistence curve in the limit $N\rightarrow
\infty $) for $T<\Theta $ we also have $h_{1}=0$ but $h_{2}\neq 0$. At the $%
\Theta $-point all three fields $h_{1},h_{2}$ and $\lambda $ vanish. In the
semi-dilute (sd) limit of polymer concentration $c_{{\rm p}}\rightarrow 0$
at fixed $\phi $ ($i.e.,$ $N\rightarrow \infty $) the osmotic pressure
becomes $\hat{\Pi}(\phi )=\phi ^{2}(\lambda +2v\phi )$ and one easily finds
from the equality of the osmotic pressure in both phases: 
\begin{equation}
\phi _{1}=0\qquad \mbox{and}\qquad \phi _{{\rm sd}}=\phi _{2}=-\frac{\lambda 
}{2v}  \label{phisd}
\end{equation}
for the polymer volume fractions in the coexisting pure solvent phase and
semi-dilute phase. Substituting this result in Eq. (\ref{h1vonphi}) we note
that we approach the symmetry plane $h_{1}=0$ for $N\rightarrow \infty $. In
the semi-dilute limit all polymer molecules aggregate in the polymer-rich
phase and their overlap $s=R_{{\rm g}}^{3}c_{{\rm p}}$ tends to infinity,
because the mean radius of gyration $R_{{\rm g}}$ diverges faster than $c_{%
{\rm p}}$ vanishes. The critical demixing point can be found from the
stability conditions 
\begin{equation}
\left( \frac{\partial \hat{\Pi}}{\partial \phi }\right) _{T,N}=\left( \frac{%
\partial ^{2}\hat{\Pi}}{\partial \phi ^{2}}\right) _{T,N}=0\,.  \label{cond2}
\end{equation}
Insertion of Eq. (\ref{exp_phi}) yields the critical volume fraction 
\begin{equation}
\phi _{{\rm c}}=-\frac{\lambda _{{\rm c}}}{6v}=\frac{1}{\sqrt{6vN}}\,.
\label{phic}
\end{equation}
>From Eqs. (\ref{phisd}) and (\ref{phic}) we can form the universal ratio $%
\phi _{{\rm sd}}/\phi _{{\rm c}}=3$, which, since $\lambda \sim T-\Theta $,
measures the ratio of the slopes of the wing critical line and the phase
separation boundary in the semi-dilute limit. The result $\phi _{{\rm sd}%
}/\phi _{{\rm c}}=3$ coincides with Flory theory but differs from the result 
$\phi _{{\rm sd}}/\phi _{{\rm c}}=5/2$ \cite{AAG:00,SF:79} that one finds
from the Landau expansion by keeping $h_{2}$ constant instead of $N$ in the
calculation of the wing critical line. To our knowledge the only attempt to
independently obtain this universal ratio was made in simulations by
Frauenkron and Grassberger \cite{FG:97,G:97}. Their result $\phi _{{\rm sd}%
}/\phi _{{\rm c}}\approx 2.9-3.2$ supports the validity of a calculation at
constant $N$.

Using the scaled variables 
\begin{eqnarray}
\Delta \tilde{\phi} &=&\frac{\phi -\phi _{{\rm c}}}{\phi _{{\rm c}}}\,,
\label{phiscal} \\
\Delta \tilde{\lambda} &=&\frac{\lambda -\lambda _{{\rm c}}}{\lambda _{{\rm c%
}}}\,,  \label{lambdascal}
\end{eqnarray}
and the mean-field result (\ref{phic}), we find that the osmotic pressure (%
\ref{exp_phi}) obeys the scaling form 
\begin{equation}
\Delta \tilde{\Pi}=\frac{\hat{\Pi}-\hat{\Pi}_{{\rm c}}}{\hat{\Pi}_{{\rm c}}}%
=-3\Delta \tilde{\lambda}(1+\Delta \tilde{\phi})^{2}+(\Delta \tilde{\phi}%
)^{3}\,,  \label{piscal}
\end{equation}
with $\hat{\Pi}_{{\rm c}}=2v\phi _{{\rm c}}^{3}$ and without explicit
system-dependent parameters. To obtain the scaling equation for the free
energy we subtract its regular part 
\begin{equation}
f_{{\rm reg}}=f(\phi _{{\rm c}})+\left. \frac{\partial f}{\partial \phi }%
\right| _{\phi _{{\rm c}}}(\phi -\phi _{{\rm c}})\,,  \label{freg}
\end{equation}
which neither does contribute to the susceptibility nor affects the
calculation of coexistence curves. With this subtraction we find 
\begin{equation}
\Delta \tilde{f}\equiv \frac{f-f_{{\rm reg}}}{\hat{\Pi}_{{\rm c}}}%
=3[(1+\Delta \tilde{\phi})\ln (1+\Delta \tilde{\phi})-\Delta \tilde{\phi}%
-\left( \Delta \tilde{\lambda}+\frac{1}{2}\right) (\Delta \tilde{\phi})^{2}+%
\frac{1}{6}(\Delta \tilde{\phi})^{3}],  \label{dfscal}
\end{equation}
and the coexistence curve is calculated from the conditions 
\begin{equation}
\left. \frac{\partial \Delta \tilde{f}}{\partial \Delta \tilde{\phi}}\right|
_{\Delta \tilde{\phi}_{1}}=\left. \frac{\partial \Delta \tilde{f}}{\partial
\Delta \tilde{\phi}}\right| _{\Delta \tilde{\phi}_{2}}=\frac{\Delta \tilde{f}%
(\Delta \tilde{\phi}_{2})-\Delta \tilde{f}(\Delta \tilde{\phi}_{1})}{\Delta 
\tilde{\phi}_{2}-\Delta \tilde{\phi}_{1}}\,.  \label{equnren}
\end{equation}
The Ansatz (\ref{lvont}) for the temperature dependence of $\lambda $ gives 
\begin{equation}
\Delta \tilde{\lambda}=\frac{T-T_{{\rm c}}}{T_{{\rm c}}-\Theta }\,,
\label{tscal}
\end{equation}
for the scaling variable $\Delta \tilde{\lambda}$. All coexistence curves
are expected to collapse onto a single scaling curve, when plotted in terms
of the variables $(T-T_{{\rm c}})/(T_{{\rm c}}-\Theta )$ and $(\phi -\phi _{%
{\rm c}})/\phi _{{\rm c}}$. Such scaling behavior was already proposed
earlier by Izumi and Miyake \cite{IM:84} based on homogeneity arguments, but
the quality of the scaling was not very good. As mentioned earlier \cite
{AAG:00}, the scaling variable (\ref{tscal}) for critical temperatures $T_{%
{\rm c}}$ close to the $\Theta $-point is rather sensitive to the value of
the $\Theta $-temperature. The extrapolation of $\Theta =\lim_{N\rightarrow
\infty }T_{{\rm c}}(N)$ is notoriously difficult due to poorly controlled
finite chain-length effects and effects of polydispersity. This is why in
the earlier work \cite{AAG:00} both $\phi -\phi _{{\rm c}}$ and $T-T_{{\rm c}%
}$ were scaled by $\phi _{{\rm c}}$. In practice, it means that $T_{{\rm c}%
}-\Theta $ was replaced by an empirical function of $\phi _{{\rm c}}$ taken
from experiment. Moreover a second-order term (quadratic in $T-T_{{\rm c}}$)
was added to account for nonasymptotic effects at lower degrees of
polymerization. Such an approach yields an almost perfect description of the
phase separation but contains too many empirical features. In this work we
try to avoid any empirical assumptions and keep $T_{{\rm c}}-\Theta $ as a
scaling factor while adjusting the value of the $\Theta $-temperature. To
derive the scaling we use the zero-loop approximation $a_{2}=\lambda $ and $%
a_{3}=2v$ for the second and third virial coefficient in the virial
expansion (\ref{exp_phi}) of the osmotic pressure. Perturbative corrections
in the two- and three-point couplings $\lambda $ and $v$ have been
calculated up to two-loop order \cite{lbook}, but since this leads to
divergent series for the virial coefficients, which one does not know how to
resum, we restrict ourselves to incorporating only the first-order
perturbation as calculated by Duplantier \cite{D:82}. The calculation to
first order in the coupling constants $\lambda $ and $v$ modifies only the
second virial coefficient, which for a monodisperse solution reads in our
notation 
\begin{equation}
a_{2}=\lambda -\frac{24v}{(2\pi )^{3/2}N^{1/2}}\,.  \label{aunren}
\end{equation}
Solving the critical point conditions (\ref{cond2}), we find that the scaled
Eq. (\ref{piscal}) for the osmotic pressure still holds, but with a modified
scaling variable 
\begin{equation}
\Delta \tilde{\lambda}=\frac{\lambda -\lambda _{{\rm c}}}{\lambda _{{\rm c}}}%
\left( 1-\frac{4}{(2\pi )^{3/2}\phi _{{\rm c}}N^{1/2}}\right) =\frac{T-T_{%
{\rm c}}}{T_{{\rm c}}-\Theta }\left( 1-\frac{4}{(2\pi )^{3/2}\phi _{{\rm c}%
}N^{1/2}}\right) \,.  \label{lambdafinscal}
\end{equation}
Inclusion of the effects of tricritical fluctuations, to be described in
Sec. \ref{trisec}, introduces a nonuniversal parameter $\rho (v_{{\rm R}%
}^{0})$ in the denominator of the correction term in Eq (\ref{lambdafinscal}%
). One can also argue, that such a factor should already be accounted for in
the unrenormalized result, since the microstructure of the real polymer
differs from that of the model underlying the calculation. Thus the relation
between the chain-length of the actual polymer and the chain-length used in
the calculations contains a nonuniversal factor. This consideration leaves
us with two fitting parameters $\Theta $ and $c=4/(2\pi \rho (v_{{\rm R}%
}^{0}))^{3/2}$, and with the chain-length $N=M_{{\rm w}}/M_{{\rm m}}$
measured as the number of monomers in a chain ($M_{{\rm m}}$ being the
monomer molecular weight).

\section{Critical Renormalization}

\label{critren}

To incorporate the influence of long-range critical fluctuations, which
strongly affect the shape of the coexistence curve in the vicinity of the
critical demixing point, we employ a procedure developed by Chen {\it et al.}
\cite{CS:90:1,CS:90:2} using renormalization-group matching \cite{NA:85} to
implement the crossover between mean field and critical behavior. The free
energy (\ref{dfscal}) can be expanded around the critical point, yielding

\begin{equation}
\Delta \tilde{f}=-3\Delta \tilde{\lambda} (\Delta \tilde{\phi} )^{2}+ \frac{1%
}{4}(\Delta \tilde{\phi} )^{4}+...\, .  \label{dfr}
\end{equation}

The renormalization-group theory for the scalar $\psi ^{4}$ field theory 
\cite{ZJ:96} provides a proper tool to account for critical fluctuations
that give rise to scale invariance at the critical point \cite{DG:76}. The
scaled free-energy of a system with order parameter $\tilde{M}$ and reduced
temperature $\tilde{t}$ in mean-field approximation can be expanded as 
\begin{equation}
\Delta \tilde{f}=\frac{1}{2}\tilde{t}\tilde{M}^{2}+\frac{u^{\ast }\bar{u}%
\tilde{\Lambda}}{4!}\tilde{M}^{4}+...\, ,  \label{dmr}
\end{equation}
where the four-point coupling $\bar{u}$ is normalized by its fixed point
value $u^{\ast }\cong 0.472$ \cite{ZJ:96,TC:91} and where $\tilde{\Lambda}$
is a cutoff length-scale. Since all system-dependent parameters are scaled
away in Eq. (\ref{dmr}), we allow for two amplitudes $\tilde{c}_{\rho }$ and 
$\tilde{c}_{t}$ in the relations 
\begin{equation}
\tilde{t}=\tilde{c}_{t}\Delta \tilde{\lambda}\,,\quad \quad \tilde{M}=\tilde{%
c}_{\rho }\Delta \tilde{\phi}\,,  \label{param}
\end{equation}
connecting our physical variables to those of the field theory. Equating the
expansion coefficients in Eqs. (\ref{dfr}) and (\ref{dmr}) we find the
relations 
\begin{equation}
\tilde{c}_{t}\tilde{c}_{\rho }^{2}=6\,,\quad \quad u^{\ast }\bar{u}\tilde{%
\Lambda}=\frac{6}{\tilde{c}_{\rho }^{4}}\,,  \label{pararestr}
\end{equation}
restricting the four parameters $\bar{u},\tilde{\Lambda},\tilde{c}_{t},%
\tilde{c}_{\rho }$ to only two, which can be varied independently. The above
expansions do not depend on the degree of polymerization. It is interesting
to compare expansions (\ref{dfr}) and (\ref{dmr}) with conventional Landau
expansions of the unscaled free energy \cite{PS:99} 
\begin{equation}
f=\frac{1}{2}a_{0}\frac{T-T_{{\rm c}}}{T_{{\rm c}}}(\Delta \phi )^{2}+\frac{1%
}{4!}u_{0}(\Delta \phi )^{4}+...\quad  \label{udfr}
\end{equation}
and with that of the reduced variables 
\begin{equation}
f=\frac{1}{2}t(M)^{2}+\frac{u^{\ast }\bar{u}\Lambda }{4!}u_{0}(M)^{4}+...\, ,
\label{udmr}
\end{equation}
where 
\begin{equation}
t=c_{t}\frac{T-T_{{\rm c}}}{T_{{\rm c}}}\,,\quad \quad M=c_{\rho }\Delta
\phi \,,  \label{uparam}
\end{equation}
and 
\begin{equation}
c_{t}c_{\rho }^{2}=a_{0}\,,\quad \quad u^{\ast }\bar{u}\Lambda =\frac{u_{0}}{%
c_{\rho }^{4}}\, .  \label{upararestr}
\end{equation}
In the Flory model we find in the large $N$ limit $a_{0}=2T_{{\rm c}}/\Theta
\lambda_{0}\rightarrow 1,u_{0}=12v/\phi _{{\rm c}} \rightarrow 2/\phi _{{\rm %
c}}$, hence $\lambda _{0}=2$ and $v=1/6$. Furthermore, since $c_{t}=c_{t0}/%
\sqrt{N},\Lambda =\Lambda _{0}/\sqrt{N}$, and $c_{\rho }=c_{\rho 0}/N^{1/4}$ 
\cite{AA:01}, comparison with Eqs. (\ref{param}) and (\ref{pararestr})
yields: $\Lambda _{0}=\tilde{\Lambda},c_{\rho 0}=\tilde{ c_{\rho }}$, and $%
c_{t0}=\tilde{c_{t}}/6$. Note, that these parameters, as well as $\bar{u}$,
do not depend on the degree of polymerization.

Renormalization proceeds by replacing the physical variables in Eq. (\ref
{dmr}) by renormalized ones defined as 
\begin{equation}
\tilde{t}\rightarrow \tilde{t}_{{\rm x}}=\tilde{t}{\cal T}\,,\quad \quad 
\tilde{M}\rightarrow \tilde{M}_{{\rm x}}=\tilde{M}{\cal D}^{1/2}\,,\quad
\quad \bar{u}\rightarrow \bar{u}_{{\rm x}}=\bar{u}{\cal U}\,,  \label{ren}
\end{equation}
where the rescaling functions ${\cal T}$, ${\cal D}$ are integrated
renormalization-group (RG) exponent functions and ${\cal U}$ is the
integrated Wilson flow function \cite{CS:90:1}. We can approximate these
functions with good accuracy in terms of a crossover variable $Y$ by \cite
{TC:91} 
\begin{equation}
{\cal T}=Y^{(2-1/\nu )/\omega },\quad {\cal D}=Y^{-\eta /\omega },\quad 
{\cal U}=Y^{1/\omega },  \label{rescaling}
\end{equation}
with $\omega =\Delta _{{\rm s}}/\nu $, where $\eta =2-\gamma /\nu =0.033\pm
0.003$, $\gamma =1.239\pm 0.002$, $\nu =0.630\pm 0.001$ , and $\Delta _{{\rm %
s}}=0.51\pm 0.02$ \cite{ZJ:96,LF:89,GZ:98,CV:99,ZF:96} are the universal
critical exponents of the asymptotic power laws for the correlation
function, susceptibility, correlation length, and for the Wegner correction
to asymptotic scaling, respectively. For the free energy itself, it is
necessary to perform an additional additive renormalization by adding the
``kernel'' term $\displaystyle-\frac{1}{2}\tilde{t}^{2}{\cal K}$ with 
\begin{equation}
{\cal K}=\frac{\nu }{\alpha \bar{u}\tilde{\Lambda}}\Bigl(Y^{-\alpha /\omega
\nu }-1\Bigr)  \label{kernel}
\end{equation}
to the free energy density. In Eq. (\ref{kernel}) $\alpha =2-3\nu =0.110\pm
0.001$ \cite{ZJ:96,LF:89,GZ:98,CV:99} is the universal critical exponent for
the heat capacity. The crossover function $Y$ is defined implicitly by the
equation 
\begin{equation}
1-(1-\bar{u})Y=\bar{u}[1+(\tilde{\Lambda}/\tilde{\kappa})^{2}]^{1/2}Y^{1/%
\omega }.  \label{Y}
\end{equation}
which evaluates the integrated flow equation for the running coupling
constant $u(l)$ at a specific matching point $l=l^{\ast }$ where one
recovers the mean-field expression for the free energy \cite{CS:90:2}. The
parameter $\tilde{\kappa}$ is defined as 
\begin{equation}
\tilde{\kappa}^{2}=\frac{\partial ^{2}\Delta \tilde{f}_{{\rm x}}}{\partial (%
{\cal D}^{1/2}\tilde{M})^{2}}{}{},  \label{kappa2}
\end{equation}
where $\Delta \tilde{f}_{{\rm x}}=\Delta \tilde{f}(\tilde{t}_{{\rm x}},%
\tilde{M}_{{\rm x}})$. It measures the distance from the critical point.
Note that the unscaled $\kappa =\tilde{\kappa}/N^{1/4}$ depends on the
degree of polymerization, being inversely proportional to the correlation
length $\xi $ \cite{AA:01,PS:99}. Asymptotically close to $T_{{\rm c}},%
\tilde{\kappa}^{2}\rightarrow 0$ and $Y\sim (\tilde{\kappa}/\bar{u}\tilde{%
\Lambda})^{\omega }$, yielding the Ising asymptotic behavior \cite{AA:01}.
Close to the $\Theta $-point $\tilde{\kappa}^{2}\sim \Delta \tilde{\lambda}%
=(T-T_{{\rm c}})/(T_{{\rm c}}-\Theta )$, diverging as $T_{{\rm c}%
}\rightarrow \Theta $ and driving the crossover function $Y$ to its
mean-field limit $Y=1$. In unscaled variables, the crossover to mean-field $%
\Theta $-point tricriticality is driven by $\Lambda \sim N^{-1/2}$, assumed
to be inversely proportional to the radius of gyration. To obtain the
crossover to mean-field tricriticality, instead of expanding the free energy 
$\Delta \tilde{f}$, we now renormalize the full expression (\ref{dfscal}) to
keep the proper low-density limit away from the Ising critical point, which
is contained in this expression. A similar approach was used earlier to
incorporate critical fluctuations into the Van der Waals equation \cite
{WA:99} and into the Flory-Huggins model \cite{PS:99}. Since Eq. (\ref
{dfscal}) contains only $\Delta \tilde{\lambda}$ and $\Delta \tilde{\phi}$
as variables, corresponding to $\tilde{t}$ and $\tilde{M}$, but no explicit
coupling $\bar{u}$, we incorporate the renormalization factor ${\cal {U}}$
into the factors renormalizing $\Delta \tilde{\lambda}$ and $\Delta \tilde{%
\phi}$ in a way, that reproduces the renormalization of (\ref{dmr}) when (%
\ref{dfscal}) is expanded. This leads to the renormalization transformations 
\begin{equation}
\Delta \tilde{\lambda}\rightarrow \Delta \tilde{\lambda}_{{\rm x}}={\cal T}%
{\cal U}^{-1/2}\Delta \tilde{\lambda}\equiv {\cal S}\Delta \tilde{\lambda}%
\,,\quad \quad \Delta \tilde{\phi}\rightarrow \Delta \tilde{\phi}_{{\rm x}}=%
{\cal D}^{1/2}{\cal U}^{1/4}\Delta \tilde{\phi}\equiv {\cal R}\Delta \tilde{%
\phi}\,,  \label{finren}
\end{equation}
and an accordingly modified definition of the crossover parameter $\tilde{%
\kappa}$ 
\begin{equation}
\tilde{\kappa}^{2}={\cal U}^{1/2}\frac{1}{\tilde{c}_{\rho }^{2}}\tilde{\chi}%
_{{\rm x}}^{-1},  \label{kappaphi}
\end{equation}
where $\tilde{\chi}_{{\rm x}}^{-1}$ is the renormalized (``crossover'')
inverse susceptibility 
\begin{equation}
\tilde{\chi}_{{\rm x}}^{-1}=\frac{\partial ^{2}\Delta \tilde{f}}{\partial
(\Delta \tilde{\phi})^{2}}(\Delta \tilde{\lambda}_{{\rm x}},\Delta \tilde{%
\phi}_{{\rm x}})=3\left( \frac{1}{1+\Delta \tilde{\phi}_{{\rm x}}}-2(\Delta 
\tilde{\lambda}_{{\rm x}}+\frac{1}{2})+\Delta \tilde{\phi}_{{\rm x}}\right) .
\label{susz}
\end{equation}
We observe that Eq. (\ref{kappaphi}) can be solved for $\Delta \tilde{\phi}_{%
{\rm x}}$ and after eliminating $\tilde{\kappa}^{2}$ with Eq. (\ref{Y}), we
obtain $\Delta \tilde{\phi}_{{\rm x}}$ as a function of the crossover
variable $Y$ and $\Delta \tilde{\lambda}$. We then numerically solve the
two-phase coexistence conditions 
\begin{equation}
\left. \frac{\partial \Delta \tilde{f}_{{\rm x}}}{\partial \Delta \tilde{\phi%
}}\right| _{\Delta \tilde{\phi}_{1}}=\left. \frac{\partial \Delta \tilde{f}_{%
{\rm x}}}{\partial \Delta \tilde{\phi}}\right| _{\Delta \tilde{\phi}_{2}}=%
\frac{\Delta \tilde{f}_{{\rm x}}(\Delta \tilde{\phi}_{{\rm x}2})-\Delta 
\tilde{f}_{{\rm x}}(\Delta \tilde{\phi}_{{\rm x}1})}{\Delta \tilde{\phi}%
_{2}-\Delta \tilde{\phi}_{1}}  \label{eqren}
\end{equation}
for $Y_{1}$ and $Y_{2}$ at a given temperature and thereby find the reduced
densities $\Delta \tilde{\phi}_{1}$ and $\Delta \tilde{\phi}_{2}$ of the
coexisting phases. The procedure outlined above implements the crossover
between the critical regime and the tricritical regime. In the critical
regime the correlation length of concentration fluctuations is larger than
the radius of gyration of a polymer, which at low enough concentration sets
the scale for tricritical fluctuations. Thus near the critical demixing
point the coexistence-curve has an Ising shape. In the tricritical regime
the radius of gyration is larger than the correlation length of
concentrationl fluctuations and the coexistence curve becomes triangle
shaped. One can quantify the location of the crossover region by a scaled
Ginzburg number, which is given in terms of our crossover parameters by 
\begin{equation}
\tilde{N}_{{\rm G}}=g_{0}\frac{(\bar{u}\tilde{\Lambda})^{2}}{\tilde{c}_{t}}%
\,,  \label{ginz}
\end{equation}
with $g_{0}\cong 0.031$ \cite{PS:99}. Fitting the theory to the experimental
data as described in Sec. \ref{comp}, we find (in agreement with
light-scattering experiments \cite{KA:02}) $\bar{u}\cong 1$ and $\tilde{c}%
_{\rho }\cong 1.6$, and with the use of Eq. (\ref{pararestr}) we obtain $%
\tilde{N}_{{\rm G}}\cong 0.05$. For $\Delta \tilde{\lambda}\ll \tilde{N}_{%
{\rm G}}$ we find Ising-type behavior and for $\Delta \tilde{\lambda}\gg 
\tilde{N}_{{\rm G}}$ we find tricritical behavior. In unscaled variables,
the actual Ginzburg number is $N_{{\rm G}}=\tilde{N_{{\rm G}}}\lambda _{{\rm %
c}}$, which in the Flory model becomes $N_{{\rm G}}\cong \tilde{N_{{\rm G}}}/%
\sqrt{N}$ \cite{PS:99}.

\section{Tricritical Renormalization}

\label{trisec} Renormalization-group treatment of the $n$-vector model in
the limit $n\longrightarrow 0$, describing tricritical behavior , reveals
that the upper critical dimension for the tricritical point is $d_{{\rm c}%
}=3 $ \cite{LS:84}. Thus, in dimension $d=3$ we expect mean-field behavior
at the tricritical point, implying that $\Theta $-point polymers on a large
scale behave effectively like random walks with a mean-field
correlation-length exponent $\nu =1/2$. In the vicinity of the tricritical
point one finds logarithmic corrections to mean-field behavior. While for
the scaled free energy (\ref{dfscal}) tricritical renormalization factors
cancel to a large extent in the scaled variables, the chain-length
dependence of the critical parameters $\phi _{{\rm c}}(N)$ and $T_{{\rm c}%
}(N)$ is modified by tricritical fluctuations. To lowest order the
tricritical renormalization-group mapping reads \cite{D:82} 
\begin{eqnarray}
\lambda &\rightarrow &\lambda _{{\rm R}}=\lambda \sigma (v_{{\rm R}}^{0})v_{%
{\rm R}}^{4/11}\,,  \label{trilamb} \\
v &\rightarrow &v_{{\rm R}}=\frac{\pi ^{2}}{33\ln N}\,,  \label{triv} \\
\phi &\rightarrow &\phi _{{\rm R}}=\phi \rho (v_{{\rm R}}^{0})(1+O(v_{{\rm R}%
}))\,,  \label{triphi} \\
N &\rightarrow &N_{{\rm R}}=N\rho (v_{{\rm R}}^{0})(1+O(v_{{\rm R}}))\,,
\label{trin}
\end{eqnarray}
where $\sigma (v_{{\rm R}}^{0})$ and $\rho (v_{{\rm R}}^{0})$ are
nonuniversal integration constants depending on the start value $v_{{\rm R}%
}^{0}$ of the renormalized three-body interaction. Renormalizing Eq. (\ref
{phic}) one finds the asymptotic relations 
\begin{equation}
\phi _{{\rm c}}\sim \frac{(\ln N)^{1/2}}{\sqrt{N}}\quad \mbox{and}\quad 
\frac{\Theta -T_{{\rm c}}}{\Theta }\sim \frac{(\ln N)^{-3/22}}{\sqrt{N}}\,.
\label{phicas}
\end{equation}
As was shown by Hager and Sch\"{a}fer \cite{HS:99}, the region where the
asymptotic relation (\ref{triv}) for $v_{{\rm R}}$ is valid is limited to
chain lengths way beyond of what is within the reach of present experiments
or simulations. Despite recent progress \cite{H:02}, the tricritical Wilson
flow function in the crossover region is still known with much lower
accuracy as compared to the critical case. We thus choose to eliminate the
running coupling $v_{{\rm R}}(N)$ in the critical-point conditions, so as to
be able to test the tricritical predictions without using some approximation
for $v_{{\rm R}}$. Without tricritical renormalization we find from Eqs. (%
\ref{phic}) and (\ref{lvont}) the mean-field relation 
\begin{equation}
\frac{\Theta -T_{{\rm c}}}{\Theta }\sim \frac{1}{\phi _{{\rm c}}N}\,.
\label{pred1}
\end{equation}
If we first renormalize Eq. (\ref{phic}) using the mapping (\ref{trilamb})-(%
\ref{trin}), we find 
\begin{equation}
\frac{\Theta -T_{{\rm c}}}{\Theta }\sim \frac{1}{\phi _{{\rm c}%
}^{3/11}N^{7/11}}\,,  \label{pred2}
\end{equation}
as the renormalized mean-field result, and if we include the first-order
correction of Eq. (\ref{aunren}) with $\lambda \rightarrow \lambda $ in
accordance with Eq. \ref{trilamb}, we obtain the relation 
\begin{equation}
\frac{\Theta -T_{{\rm c}}}{\Theta }\sim \frac{1}{\phi _{{\rm c}%
}^{3/11}N^{7/11}}\left( 1-\frac{4}{(2\pi \rho (v_{{\rm R}}^{0}))^{3/2}\phi _{%
{\rm c}}N^{1/2}}\right) \,.  \label{pred3}
\end{equation}
Besides the $\Theta $-temperature and the parameter $\rho (v_{{\rm R}}^{0})$%
, these relations contain only directly measurable quantities and we shall
test their validity in the next section. Let us now obtain the chain-length
dependent Ginzburg number $N_{{\rm G}}(N)=\tilde{N}_{{\rm G}}\lambda _{{\rm c%
}}$. With the use of the asymptotic tricritical result (\ref{phicas}) we
find 
\begin{equation}
N_{{\rm G}}(N)=\tilde{N}_{{\rm G}}\lambda _{{\rm c}}\sim \frac{\ln ^{-3/11}N%
}{\sqrt{N}}\,.  \label{nginz}
\end{equation}
Thus the Ising region shrinks to zero, when we approach the $\Theta $-point $%
(N\rightarrow \infty )$ but the classical $N^{-1/2}$ dependence is now
modified by tricritical fluctuations. By comparing the expansion
coefficients of the unscaled free energy $f-f_{{\rm reg}}$ to Eq. (\ref{dmr}%
) and with the use of Eq. (\ref{ginz}) and the asymptotic tricritical
results (\ref{phicas}), we find the chain-length dependent parameters 
\begin{eqnarray}
\bar{u}(N)\Lambda (N) &=&\phi _{{\rm c}}v^{1/3}\left( \frac{12u^{\ast }%
\tilde{N}_{{\rm G}}^{2}}{g_{0}^{2}}\right) ^{1/3}\sim \frac{\ln ^{1/6}N}{%
\sqrt{N}}\,, \\
c_{{\rm t}}(N) &=&\frac{\phi _{{\rm c}}^{2}}{\lambda _{{\rm c}}}%
v^{2/3}\left( \frac{144u^{\ast }{}^{2}\tilde{N}_{{\rm G}}}{g_{0}}\right)
^{1/3}\sim \frac{\ln ^{31/66}N}{\sqrt{N}}\,, \\
c_{{\rm \rho }}(N) &=&\phi _{{\rm c}}^{-1/2}v^{1/6}\left( \frac{12g_{0}}{%
u^{\ast }{}^{2}\tilde{N}_{{\rm G}}}\right) ^{1/6}\sim \sqrt{N}\ln
^{-1/12}N\,.
\end{eqnarray}
Apart from the logarithmic terms, due to the tricritical renormalization,
the chain-length dependence is the same as that for the Flory model \cite
{PS:99}.

\section{Comparison with Experimenal and Simulation Data}

\label{comp}

\subsection{Experimental coexistence curves}

\label{scalfit}

We first analyze three sets of experimental coexistence-curve data available
in the literature \cite{NC:78,NK:75,KK:75,DK:80,XS:96,XW:92}. The original
data are displayed in Figs. 2-4. The first set of data is on the system
polystyrene (PS) in cyclohexane \cite{NC:78,NK:75,KK:75}, with coexistence
curves for three different molecular weights $M_{{\rm w}%
}=110,000,200,000,1,560,000$. The next set contains data for seven different
chain lengths for the system PS in methylcyclohexane \cite{DK:80} with
molecular weights ranging from $10,200-719,000$. Another data set covers six
chain lengths for the system polymethylmethacrylate (PMMA) in $3$-octanone 
\cite{XS:96,XW:92}, with molecular weights ranging from $26,900-596,000$. In
all cases the polymer fractions were reasonably monodisperse with
polydispersity indices in the range $M_{{\rm w}}/M_{{\rm n}}=1.03-1.11$,
where $M_{{\rm w}}$ and $M_{{\rm n}}$ are the weight averaged and number
averaged molecular weight, respectively. For the PS data we used all data
points as published. For the PMMA data we used the critical point volume
fraction $\phi _{{\rm c}}=(\phi _{1}+\phi _{2})/2$ with $\phi _{1}$ and $%
\phi _{2}$ being the pair of coexisting densities closest to the critical
point. This leads to critical volume fractions being on the average $4\%$
smaller than the published ones. As a first step we fitted the parameters $%
\Theta $ and $\rho (v_{{\rm R}}^{0})$, contained in the scaling variable $%
\Delta \tilde{\lambda}$ given by Eq. (\ref{lambdafinscal}), to find an
optimal collapse for the data within each data set when, plotted in terms of
the scaling variables $\Delta \tilde{\lambda}$ and $\Delta \tilde{\phi}$. It
turns out, that $\rho (v_{{\rm R}}^{0})=1$ is a reasonable choice for all
three data sets. This leads to the coefficient $4/(2\pi )^{3/2}=4/[2\pi \rho
(v_{{\rm R}}^{0})]^{3/2}=0.254$ in the correction term of Eq. (\ref
{lambdafinscal}) and, consequently, in Eq. (\ref{pred3}). The contribution
from the correction term in Eq. (\ref{lambdafinscal}) is about $13\%$ or
smaller for all data sets. The fitted $\Theta $-temperatures are: $\Theta
=309$ K for PS in cyclohexane, $\Theta =345$ K for PS in methylcyclohexane
and $\Theta =345.65$ K for PMMA in $3$-octanone. Our value for PMMA in $3$%
-octanone lies between the values $\Theta =345.15$ K \cite{XW:92} and $%
\Theta =346.85$ K \cite{XS:96}, obtained by extrapolating $1/T_{{\rm c}}(N)$%
, taking $N^{-1/2}$ corrections into account. Our estimates for PS in
cyclohexane and in methylcyclohexane lie slightly above the values $\Theta
=307.15$ K and $\Theta =341.95$ K obtained by Izumi and Miyake \cite{IM:84},
who used the mean-field scaling variable (\ref{tscal}). The difference is
due to the correction term in Eq. (\ref{lambdafinscal}), which leads to a
better data collapse than in Ref. \cite{IM:84}, but for slightly higher $%
\Theta $-temperatures. 

\begin{figure}[-t] \begin{center}
\epsfig{figure=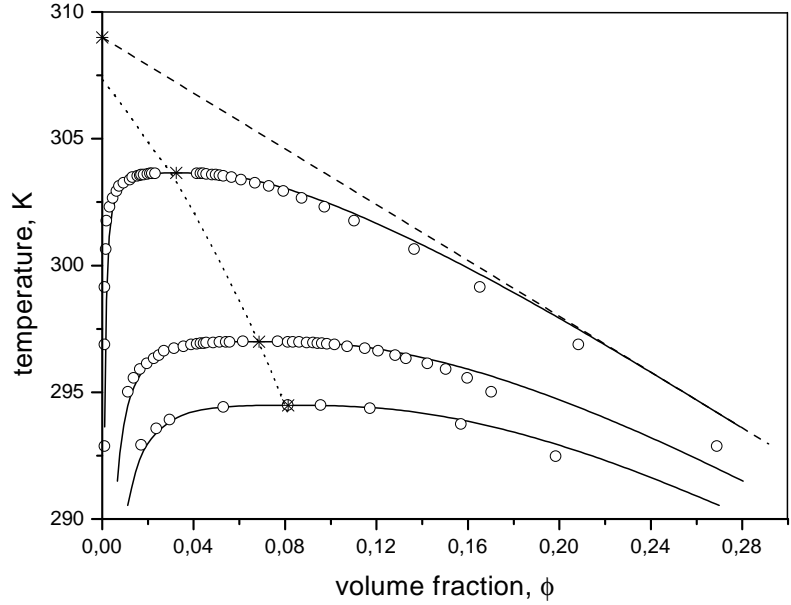,width=8cm,angle=270} 
\vspace{0.5cm}
\caption[experimental psc]{Coexistence-curve data for PS in cyclohexane 
\protect\cite{NC:78,NK:75,KK:75} are shown together with the result of the
renormalized crossover theory (full curves). The dotted curve represents Eq.
(\ref{pred3}) with $\Theta =307.25$ K. The critical demixing points and the
coexistence-curve scaling value of $\Theta =309$ K are denoted by stars. An
estimate for the limiting phase-separation boundary with $\protect\phi _{%
{\rm sd}}/\protect\phi _{{\rm c}}=3$ is shown by the dashed line.}
\label{fig4}
\end{center} 
\end{figure}
\begin{figure}[-t] \begin{center}
\epsfig{figure=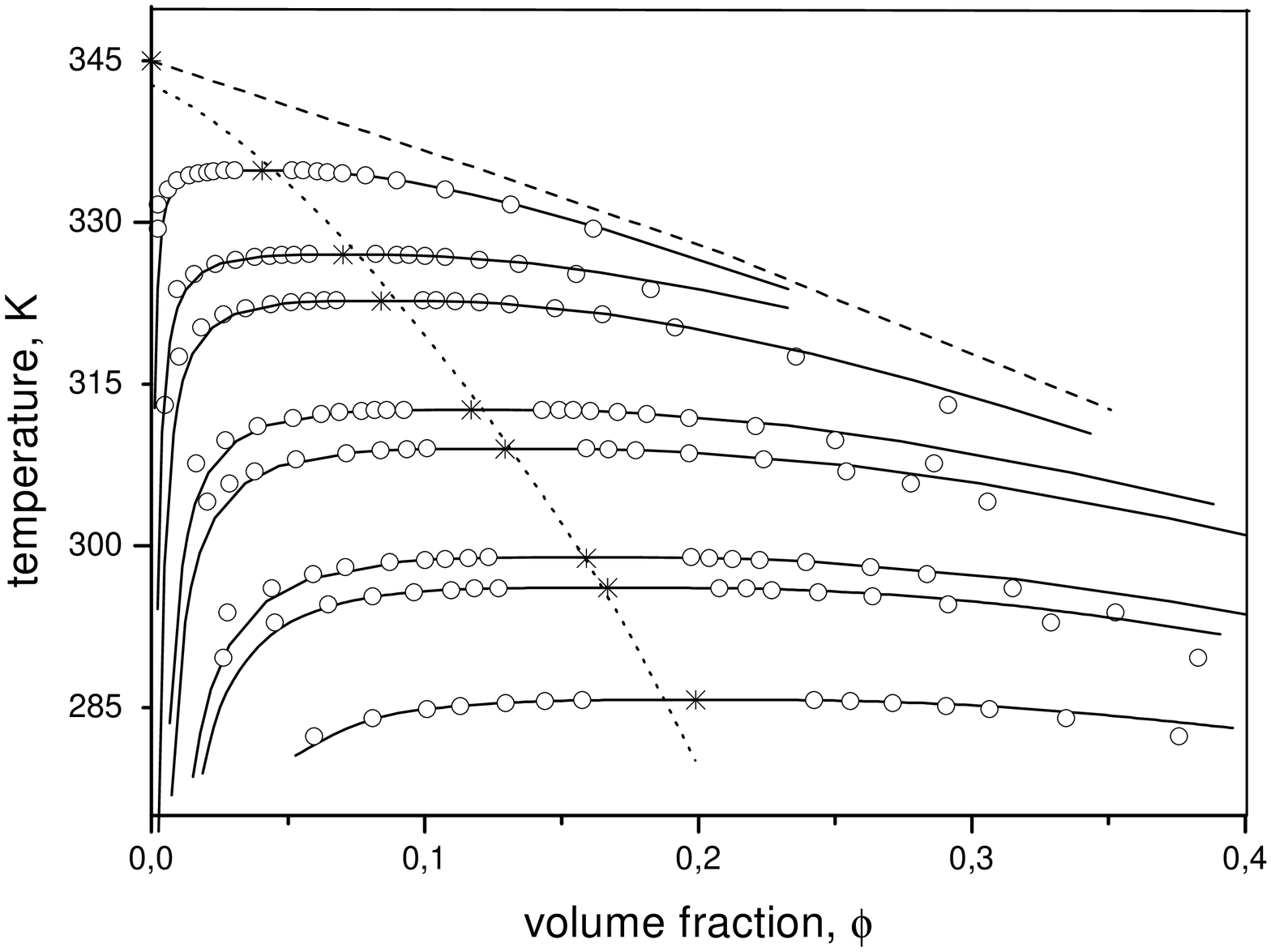,width=8cm,angle=270} 
\vspace{0.5cm}
\caption[experimental psmc]{Coexistence-curve data for PS in
methylcyclohexane \protect\cite{DK:80} are shown together with the result of
the renormalized crossover theory (full curves). The dotted curve represents
Eq. (\ref{pred3}) with $\Theta =342.75$ K. The critical demixing points and
the coexistence-curve scaling value of $\Theta =345$ K are denoted by stars.
An estimate for the limiting phase-separation boundary with $\protect\phi _{%
{\rm sd}}/\protect\phi _{{\rm c}}=3$ is shown by the dashed line.}
\label{fig5}
\end{center} 
\end{figure}
\begin{figure}[-t] \begin{center}
\epsfig{figure=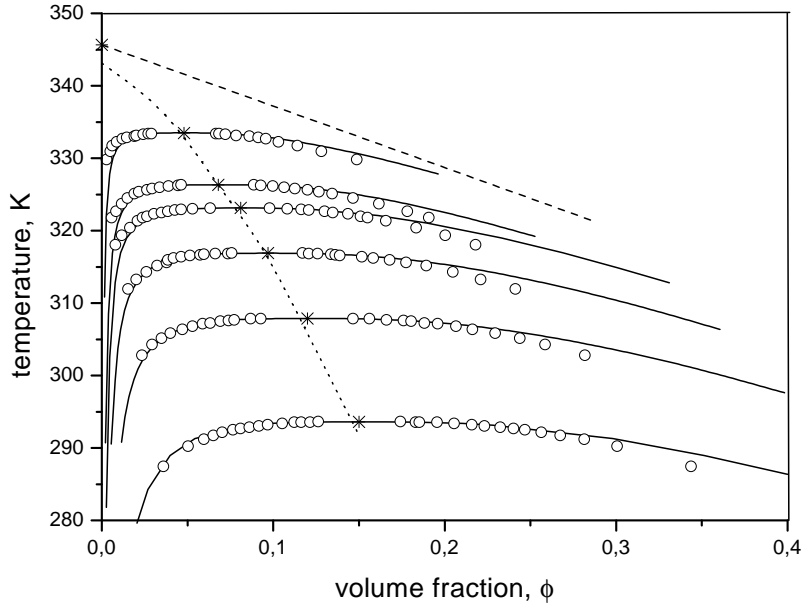,width=8cm,angle=270} 
\vspace{0.5cm}
\caption[experimental pmma]{Coexistence-curve data for PMMA in $3$-octanone 
\protect\cite{XS:96,XW:92} are shown together with the result of the
renormalized crossover theory. The dotted curve represents Eq. (\ref{pred3})
with $\Theta =343.15$ K. The critical demixing points and the
coexistence-curve scaling value of $\Theta =345.65$ K are denoted by stars.
An estimate for the limiting phase-separation boundary with $\protect\phi _{%
{\rm sd}}/\protect\phi _{{\rm c}}=3$ is shown by the dashed line.}
\label{fig6}
\end{center} 
\end{figure}

Compared to other methods, that use mean-field
extrapolations of the critical parameters $\phi _{{\rm c}}$ and $T_{{\rm c}}$
leading to values $\Theta =305.6$ - $308.4$ K (see Ref. \cite{KD:00} p. 163)
for the system PS in cyclohexane, our value is at the upper end of the
estimated range. This method of fitting the $\Theta $-temperature has the
advantage, that no explicit $N$-dependence for $T_{{\rm c}}(N)$ needs to be
assumed. The scaled data are displayed on a linear scale in Fig. \ref{fig5}
and in a logarithmic scale in Fig. \ref{fig6}. Note that the one data set of
Ref. \cite{XS:96} with $M_{{\rm w}}=26,900,$ which is significantly off the
scaling curve (see log plot), has also the largest polydispersity index $M_{%
{\rm w}}/M_{{\rm n}}=1.11$ of all samples. It is interesting to observe that
the data for all three systems with good accuracy collapse onto a single
scaling curve, as is suggested by the solution of the unrenormalized
coexistence conditions (\ref{equnren}) indicated by a dashed line in Figs. 
\ref{fig5} and \ref{fig6}. This is not necessarily to be expected, since the
critical renormalization introduces two additional parameters $\bar{u}$ and $%
\tilde{c}_{\rho }$ which may have different values for different solutes and
solvents. One clearly sees, that the unrenormalized scaled coexistence curve
fails to reproduce the proper Ising-type singularity (\ref{ising}) with $%
\beta =0.325$ close to the critical point, but instead has the mean-field
exponent $\beta =1/2$. In our second step in fitting the data we remedy this
deficiency by applying the renormalization procedure as outlined in Sec. \ref
{critren}. Since the parameter $\bar{u}$ was already found to be close to
unity in a recent evaluation of light-scattering data above $T_{{\rm c}}$ 
\cite{KA:02}, and its variation does not affect the coexistence curves very
much, we fixed it to $\bar{u}=1$, leaving $\tilde{c}_{\rho }$ as the only
fit parameter of the crossover theory. The result of our fit with $\tilde{c}%
_{\rho }=1.6$ is displayed in Figs. 2-4 in terms of unscaled variables and
in Figs. 5 and 6 in terms of scaled variables. Our scaled crossover
formulation nicely reproduces the Ising singularity, but shows some
deviations in the crossover region which cannot be removed by tuning $\tilde{%
c}_{\rho }$. One can think of a plethora of possible sources for such a
deviation, since in our approach we neglected higher-order terms in
expansions at several stages in our calculations. We truncated the virial
expansion at order $\phi ^{3}$, included only first-order perturbative
corrections in $\lambda $ and $v$, and used the lowest-order approximations
for the temperature dependence of $\lambda $ and $v$. The influence of
higher-order terms in the expansion of the temperature dependence of $%
\lambda $ and $v$ and similarly of higher-order terms in the perturbation
theory causes only a change of the scaling variable $\Delta \tilde{\lambda}$%
, that cannot account for the deviation. Note that both coexisting densities 
$\phi _{1}$ and $\phi _{2}$ of the renormalized fit are shifted to higher
densities as compared to the experimental data. Higher-order terms in the
virial expansion, like a negative $\phi ^{4}$-term can induce a shift of
both coexisting densities in the double-tangent construction (\ref{eqren})
to lower values. Since we use the theory to volume fractions up to $\phi
\leq 0.4$, it is likely that such higher-order terms in the virial expansion
contribute to the observed deviations.

\begin{figure}[-t] \begin{center}
\epsfig{figure=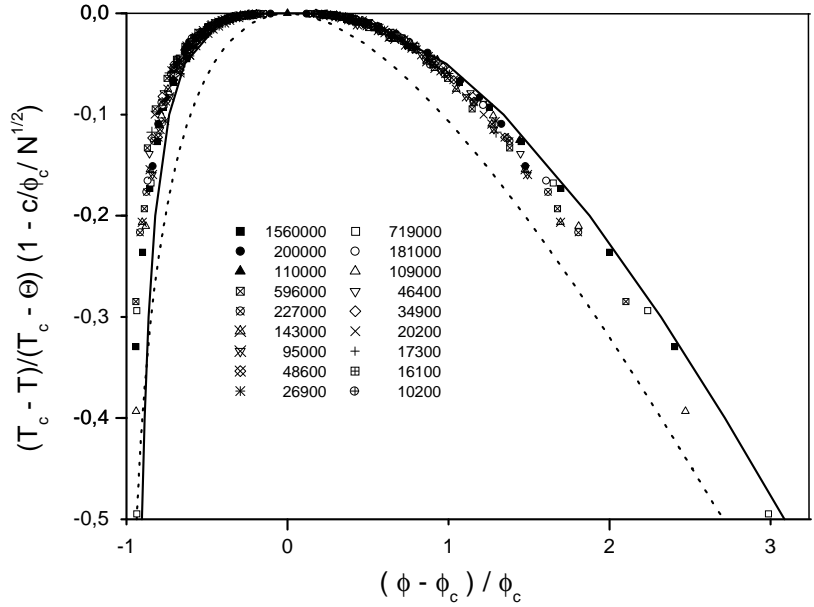,width=8cm,angle=270} 
\vspace{0.5cm}
\caption[Linear experimental scaling]{Linear scaling plot of
phase-coexistence curves for PS in cyclohexane \protect\cite
{NC:78,NK:75,KK:75} (full symbols), PS in methylcyclohexane \protect\cite
{DK:80} (open symbols) and PMMA in $3$-octanone \protect\cite{XS:96,XW:92}
(open crossed symbols). The dashed curve represents the values calculated
from the unrenormalized scaled free energy (\ref{dfscal}) and the full line
represents the result of renormalized crossover theory with $\bar{u}=1$ and $%
\tilde{c}_{\protect\rho }=1.6$.}
\label{fig2}
\end{center} 
\end{figure}
\begin{figure}[-t] \begin{center}
\epsfig{figure=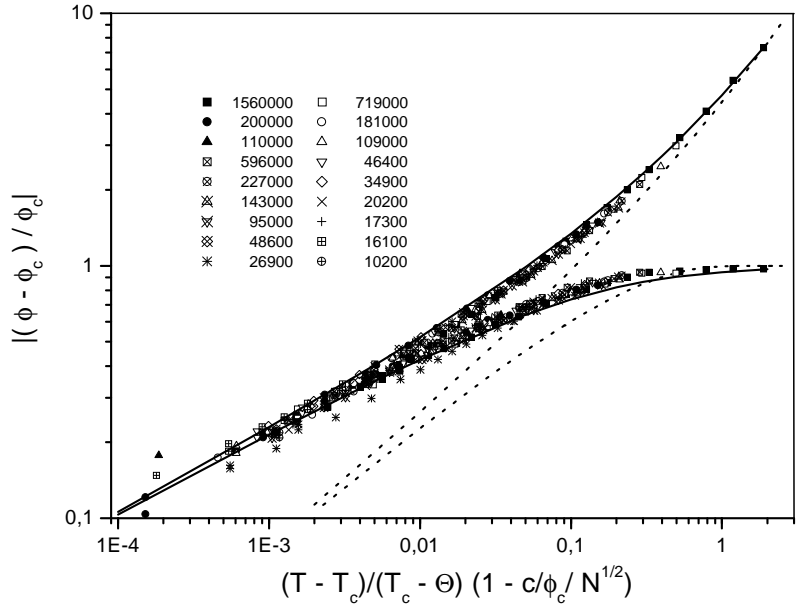,width=8cm,angle=270} 
\vspace{0.5cm}
\caption[Logarithmic experimental scaling]{Logarithmic scaling plot of
phase-coexistence curves for PS in cyclohexane \protect\cite
{NC:78,NK:75,KK:75} (full symbols), PS in methylcyclohexane \protect\cite
{DK:80} (open symbols) and PMMA in $3$-octanone \protect\cite{XS:96,XW:92}
(open crossed symbols). The dashed curve represents the values calculated
from the unrenormalized scaled free energy (\ref{dfscal}) and the full line
represents the result of renormalized crossover theory with $\bar{u}=1$ and $%
\tilde{c}_{\protect\rho }=1.6$.}
\label{fig3}
\end{center} 
\end{figure}

\subsection{Scaling of the critical parameters}

\begin{figure}[-t] \begin{center}
\epsfig{figure=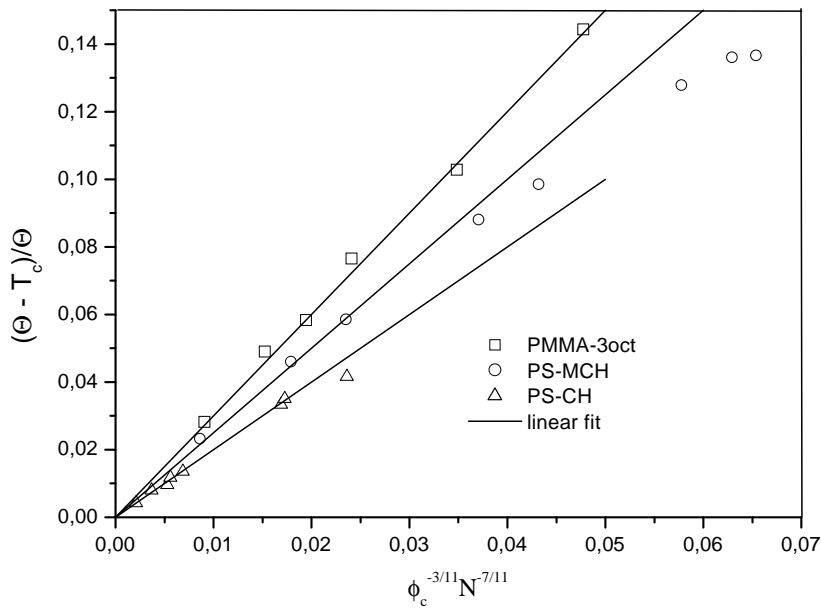,width=8cm,angle=270} 
\vspace{0.5cm}
\caption[kritfit]{The reduced temperature difference between the critical
point and the $\Theta $-point is plotted versus $\protect\phi _{{\rm c}%
}^{-3/11}N^{-7/11}$. The symbols denote the critical parameter data for
three experimental systems. The lines indicate the asymptotic linear
approximation in accordance with Eq. (\ref{pred2}).}
\label{fig10}
\end{center} 
\end{figure}

\begin{figure}[-t] \begin{center}
\epsfig{figure=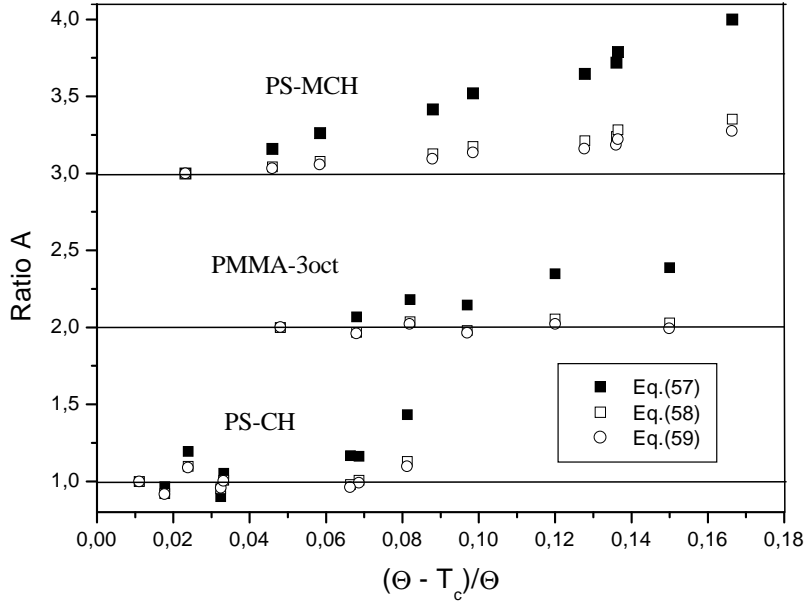,width=8cm,angle=270} 
\vspace{0.5cm}
\caption[ratio A]{The ratio $A$ (right side / left side of Eqs. (\ref{pred1}%
)-(\ref{pred3})) plotted against the reduced difference between the critical
temperature and the $\Theta $-temperature. The experimental data for three
different systems are separated by an offset of magnitude $1$.}
\label{fig9}
\end{center} 
\end{figure}

In the previous section we assigned values for the $\Theta $-temperature
that optimized scaling of the coexistence curves. An alternative procedure
to estimate the $\Theta $-temperatures is by extrapolating $N\rightarrow
\infty $ in accordance with Eqs. (\ref{pred2}) and (\ref{pred3}). The values
thus obtained for $(\Theta -T_{{\rm c}})/\Theta $ are plotted in Fig. \ref
{fig10} as a function of $\phi _{{\rm c}}^{-3/11}N^{-7/11}$. The solid lines
indicate the asymptotic linear relations in accordance with Eq. (\ref{pred2}%
). The slope of this linear relation was found to be $2$ for PS in
cyclohexane, $2.5$ for PS in methylcyclohexane and $3.5$ for PMMA in $3$%
-octanone. The resulting estimates for the $\Theta $-temperatures are $%
\Theta =307.25$ K for PS in cyclohexane, $\Theta =342.75$ K for PS in
methylcyclohexane and $\Theta =343.15$ K for PMMA in $3$-octanone. They all
lie about $0.7\%$ below the values found from coexistence-curve scaling.
Those for polystyrene compare well with other extrapolations of
critical-point data \cite{KD:00}. We believe that the origin of the
discrepancy again lies in the neglect of higher-order terms in our
calculations. To discriminate between the relations (\ref{pred1}) - (\ref
{pred3}) we form a ratio $A$, by dividing the right-hand side of each
equation by its left-hand side. The resulting ratio should be constant. The
values of the ratio $A$ for Eqs. (\ref{pred1}) - (\ref{pred3}) and the three
data sets are displayed in Fig. \ref{fig9}, with the left-most values being
normalized to unity within each set and with an additive offset to separate
the data sets. One clearly sees, that the unrenormalized prediction of Eq. (%
\ref{pred1}) gives the worst fit and this cannot be improved by lowering the 
$\Theta $-temperature, since then the curves start bending upwards close to $%
\Theta $-temperature. The renormalized fits for PS in cyclohexane and PMMA
in $3$-octanone are indeed constant within the scattering of the data, but
the most precise data on PS in methylcyclohexane clearly show a residual
slope which can be easily accounted for by introducing a correction $(\Theta
-T_{{\rm c}})^{2}$. This, together with the mismatch to the $\Theta $%
-temperatures from coexistence-curve scaling, indicates the presence of
second-order terms. Our fits for the critical-point scaling are included as
dotted lines in Figs. \ref{fig2}-\ref{fig4}. We also included the $\Theta $%
-temperatures obtained from coexistence-curve scaling and a guess for the
limiting phase boundary $\phi _{{\rm sd}}$, assuming the ratio $\phi _{{\rm %
sd}}/\phi _{{\rm c}}=3$. Note that a value of $\phi _{{\rm sd}}/\phi _{{\rm c%
}}=5/2$ is not in accord with the experimental data unless one assumes the
presence of large corrections to the asymptotic value. We think this is
unlikely since the value of the ratio is not altered by the leading order of
tricritical renormalization \cite{FG:97}.

\begin{figure}[-t] \begin{center}
\epsfig{figure=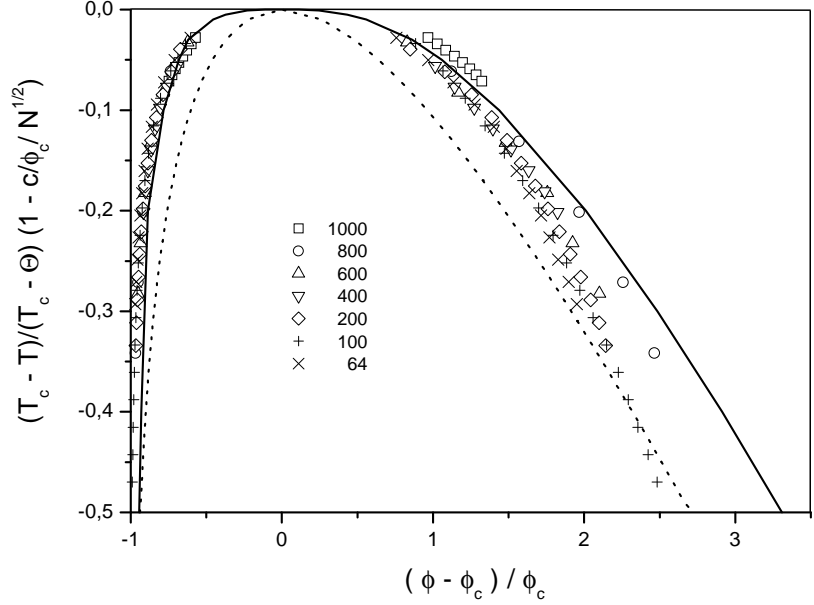,width=8cm,angle=270} 
\vspace{0.5cm}
\caption[Linear simulational scaling]{Linear scaling plot of simulated
phase-coexistence curves \protect\cite{PW:98} for self-avoiding walks with
nearest-neighbor interaction on a simple cubic lattice. The dotted curve
represents the values calculated from the unrenormalized scaled free energy (%
\ref{dfscal}) and the solid curve represents the renormalized crossover
theory with $\bar{u}=1$ and $\tilde{c}_{\protect\rho }=1.5$.}
\label{fig7}
\end{center} 
\end{figure}

\subsection{Simulations}

\begin{figure}[-t] \begin{center}
\epsfig{figure=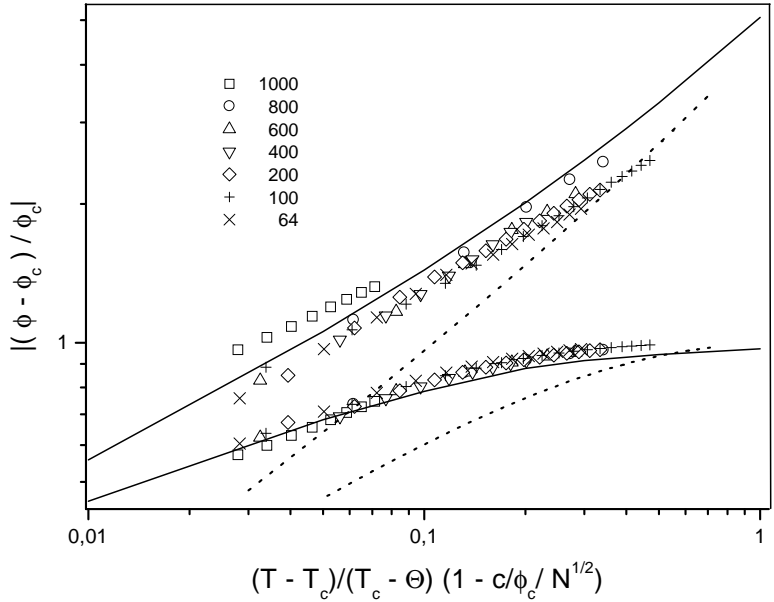,width=8cm,angle=270} 
\vspace{0.5cm}
\caption[Logarithmic simulational scaling]{Logarithmic scaling plot of
simulated phase-coexistence curves \protect\cite{PW:98} for self-avoiding
walks with nearest-neighbor interaction on a simple cubic lattice. The
dotted curve represents the values calculated from the unrenormalized scaled
free energy (\ref{dfscal}) and the solid curve represents the renormalized
crossover theory with $\bar{u}=1$ and $\tilde{c}_{\protect\rho }=1.5$.}
\label{fig8}
\end{center} 
\end{figure}

We have also analyzed a set of coexistence-curve data obtained from computer
simulations of self-avoiding walks with attractive nearest-neighbor
interaction on a simple cubic lattice \cite{PW:98} along the same lines as\
was done for the experimental coexistence-curve data in Sec. \ref{scalfit}.
We considered seven sets of coexistence-curve data with chain lengths
ranging from $64$ to $1,000$ monomers. Figure \ref{fig7} shows a linear
scaling plot of the simulation data and Fig. \ref{fig8} a logarithmic one.
The optimal data collapse was obtained for $\Theta =3.754$, measured in
units of the nearest-neighbor interaction energy, and $c=0.51$. The value
for $\Theta $ again is $1\%$ higher than the best value $\Theta =3.717(2)$
obtained from extrapolation of $1/T_{{\rm c}}$ \cite{FG:97}. In general the
quality of the data collapse is less satisfactory, compared to the
experimental data. Especially on the semi-dilute branch the data for $N=800$
and $N=1,000$ deviate significantly from those for smaller chain lengths.
One main difference between experiment and simulations is that the
simulations were done in rather small boxes. These boxes were chosen to be
at least four times larger than the maximum value of the radius of gyration.
However, with the mean squared end-end distance of a random walk being a
factor $\sqrt{6}$ larger, this still implies that a single polymer stretches
over a significant portion of the box, causing finite-size effects. Also the
rather small and varying total number of polymers, ranging between $80$ and $%
350$ for different simulations may cause problems in obtaining reliable bulk
estimates, since large surface contributions are to be expected. On the
other hand, we observe a clear breakdown of coexistence-curve scaling for
the simulation data for the shorter chain lengths $8,16$ and $32$. This is
to be expected, since we then reach volume fractions where our virial
expansion is no longer accurate. Furthermore, the symmetric coexistence
curve of the Ising model for $N=1$ cannot be scaled onto a asymmetric
scaling curve with a finite reparametrization. The two fit parameters $\bar{u%
}$ and $\tilde{c}_{\rho }$ of our crossover theory are not sufficient to
provide a decent fit to the scaling curve, probably due to the reasons
already discussed for the fit of the experimental data. More work has to be
invested in order to decide, whether the observed mismatch is due to a
limitation of the coexistence-curve scaling or can be attributed to
systematic deviations in the simulation data.

\section{Conclusion}

\label{concl} We find that the scaling description for coexistence curves of
polymer solutions, including first-order perturbative corrections, provides
a decent collapse of the available experimental data onto a single scaling
curve at fitted $\Theta $-temperatures which are systematically about $0.7\%$
higher than those obtained from an extrapolation of critical point data to
infinite chain-length. Incorporation of the critical fluctuations via
crossover RG theory leads to a decent description of the scaled coexistence
curve by fitting essentially only one parameter $\tilde{c}_{\rho }=1.6$. We
attribute the remaining deviations in the fit and the discrepancies in the $%
\Theta $-temperature estimates mainly to two approximations in our approach.
First, we truncated the virial expansion after the $\phi ^{3}$ term,
neglecting higher-order terms that contribute at larger volume fractions,
and will perturb the scaling description as they do in the Flory-Huggins
model. Another source of corrections are higher-order terms in the
temperature dependence of the bare parameters and higher-order corrections
of perturbation theory, which have similar effects on the scaling variable $%
\Delta \tilde{\lambda}$. We find that both simulations and experiments are
pointing towards an universal ratio $\phi _{{\rm sd}}/\phi _{{\rm c}}=3$,
implying that $\phi _{{\rm c}}$ has to be calculated at constant chain
length $N$ rather than at constant field $h_{2}$, which would lead to $\phi
_{{\rm sd}}/\phi _{{\rm c}}=5/2$ in the Landau theory. The scaling
description is less satisfactory for coexistence-curve data provided by
simulations. Concerning the $\Theta $-temperature extrapolations we find
from the simulation data the same pattern as for the experimental data, with
the value of $\Theta $, obtained from the scaling description, being $1\%$
higher than the best value obtained from a extrapolation of $1/T_{{\rm c}%
}(N) $.

\subsection*{Acknowledgments}

The authors acknowledge a fruitful collaboration with V.A. Agayan in an
earlier stage of this research \cite{AAG:00} and stimulating interactions
with A. Z. Panagiotopoulos and B. Widom. The research has been supported by
the Chemical Sciences, Geosciences and Biosciences Division, Office of Basic
Energy Sciences, Office of Science, U.S. Department of Energy under Grant
No. DE-FG-02-95ER-14509.


\begin{references}
\bibitem{KS:84}  C. M. Knobler and R. L. Scott, in: {\it Phase Transitions
and Critical Phenomena}, Vol. 9, edited by C. Domb and J. L. Lebowitz,
(Academic Press, New York, 1984), p. 163.

\bibitem{A:91}  M. A. Anisimov, {\it Critical Phenomena in Liquids and
Liquid Crystals}, (Gordon and Breach, Philadelphia, 1991).

\bibitem{deGennes}  P. G. de Gennes, {\ }{\it Scaling Concepts in Polymer
Physics}, (Cornell University Press, Ithaca, NY, 1979).

\bibitem{dG:78}  P. G. de Gennes, {J. Phys. (Paris)} {\bf 39}, L299 (1978).

\bibitem{dG:75}  P. G. de Gennes, {J. Phys. (Paris)} {\bf 36}, L55 (1975).

\bibitem{F:94}  M. E. Fisher, {J. Stat. Phys.} {\bf 75}, 1 (1994).

\bibitem{desC:75}  J. des Cloizeaux, {\ J. Phys. (Paris)} {\bf 36}, 281
(1975).

\bibitem{M:77}  M. A. Moore, {\ J. Phys. (Paris)} {\bf 38}, 265 (1977).

\bibitem{SW:77}  L. Sch{\"{a}}fer and T. A. Witten, {\ J. Chem. Phys.} {\bf %
66}, 2121 (1977).

\bibitem{WP:81}  J. C. Wheeler and P. Pfeuty, {J. Chem. Phys.} {\bf 74},
6415 (1981).

\bibitem{S:75}  M. J. Stephen, {Phys. Lett. A.} {\bf 53}, 363 (1975).

\bibitem{D:82}  B. Duplantier, {J. Phys. (Paris)} {\bf 43}, 991 (1982).

\bibitem{HS:99}  J. Hager and L. Sch{\"{a}}fer, {Phys. Rev. E} {\bf 60},
2071 (1999).

\bibitem{Fl:53}  P. J. Flory, {\it Principles of Polymer Chemistry},
(Cornell University Press, Ithaca, NY, 1953), Ch. XIII., M. Huggins, J.
Phys. Chem. {\bf 46}, 151 (1942).

\bibitem{DO:96}  M. Doi, {\it Introduction to Polymer Physics}, (Clarendon
Press, Oxford, 1996).

\bibitem{Wi:93}  B. Widom, {Physica A} {\bf 194}, 532 (1993).

\bibitem{MH:97}  Y. B. Melnichenko, M. A. Anisimov, A. A. Povodyrev, G. D.
Wignall, J. V. Sengers, and W. A. Van Hook, {Phys. Rev. Lett.} {\bf 79},
5266 (1997).

\bibitem{KA:02}  M. A. Anisimov, A. Kostko, and J. V. Sengers, Phys. Rev. E 
{\bf 65}, 051805 (2002).

\bibitem{AA:01}  V. A. Agayan , M. A. Anisimov, and J. V. Sengers, {Phys.
Rev. E }{\bf 64}, 026125 (2001).

\bibitem{PS:99}  A. A. Povodyrev, M.A. Anisimov, and J.V. Sengers, {Physica A%
} {\bf 264}, 345 (1999).

\bibitem{CS:90:1}  Z. Y. Chen, P.C. Albright, and J. V. Sengers, {Phys. Rev.
A} {\bf 41}, 3161 (1990).

\bibitem{CS:90:2}  Z. Y. Chen, A. Abbaci, S. Tang, and J. V. Sengers, {Phys.
Rev. A} {\bf 42}, 4470 (1990).

\bibitem{SW:80}  L. Sch{\"{a}}fer and T. A. Witten, {J. Phys. (Paris)} {\bf %
41}, 459 (1980).

\bibitem{lbook}  L. Sch{\"{a}}fer, {\ }{\it Excluded Volume Effects in
Polymer Solutions}, (Springer, Heidelberg, 1999).

\bibitem{AAG:00}  M. A. Anisimov, V. A. Agayan and E. E. Gorodetskii, {JETP
Lett.} {\bf 72}, 578 (2000).

\bibitem{SF:79}  S. Sarbach and M. E. Fisher, {Phys. Rev. B} {\bf 20}, 2797
(1979).

\bibitem{FG:97}  H. Frauenkron and P. Grassberger, {J. Chem. Phys.} {\bf 107}%
, 9599 (1997).

\bibitem{G:97}  P. Grassberger, {Phys. Rev. E} {\bf 56}, 3682 (1997).

\bibitem{NA:85}  J. F. Nicoll and P. C. Albright, {Phys. Rev. B} {\bf 31,}
4576 (1985).

\bibitem{ZJ:96}  J. Zinn-Justin, {\it Quantum Field Theory and Critical
Phenomena}, (Clarendon Press, Oxford, 1996).

\bibitem{DG:76}  C. Domb and M. S. Green (eds.), {\it Phase Transitions and
Critical Phenomena}, Vol. 6 (Academic Press, New York, 1976).

\bibitem{TC:91}  S. Tang, J. V. Sengers, and Z. Y. Chen, {Physica A} {\bf %
179,} 344 (1991).

\bibitem{LF:89}  A. J. Liu and M. E. Fisher, {Physica A} {\bf 156,} 35
(1989).

\bibitem{GZ:98}  R. Guida and J. Zinn-Justin, {J. Phys. A} {\bf 31,} 8103
(1998).

\bibitem{CV:99}  M. Campostrini, A. Pelisettimo, P. Rossi and E. Vicari, {%
Phys. Rev. E} {\bf 60,} 3526 (1999).

\bibitem{ZF:96}  S.-Y. Zinn and M. E. Fisher, {Physica A} {\bf 226,} 168
(1996).

\bibitem{IM:84}  Y. Izumi and Y. Miyake, {J. Chem. Phys.} {\bf 81}, 1501
(1984).

\bibitem{Sa:89}  I. C. Sanchez, {J. Phys. Chem.} {\bf 93}, 6983 (1989).

\bibitem{WA:99}  A. Kostrowicka Wyczalkowska, M.A. Anisimov, and J.V.
Sengers, {Fluid Phase Equilibria} {\bf 158-160}, 523 (1999).

\bibitem{LS:84}  I. D. Lawrie and S. Sarbach, in: {\it Phase Transitions and
Critical Phenomena}, Vol. 9, edited by C. Domb and J. L. Lebowitz (Academic
Press, New York, 1984), p. 1.

\bibitem{H:02}  J. S. Hager, {J. Phys. A} {\bf 35}, 2703 (2002).

\bibitem{NC:78}  M. Nakata, T. Dobashi, N. Kuwahara, M. Kaneko, and B. Chu, {%
Phys. Rev. A} {\bf 18}, 2683 (1978).

\bibitem{NK:75}  M. Nakata, N. Kuwahara, and M. Kaneko, {J. Chem. Phys.} 
{\bf 62}, 4278 (1975).

\bibitem{KK:75}  J. Kojima, N. Kuwahara, and M. Kaneko, {J. Chem. Phys.} 
{\bf 63}, 333 (1975).

\bibitem{DK:80}  T. Dobashi, M. Nakata, and M. Kaneko, {J. Chem. Phys.} {\bf %
72}, 6685 (1980); {\it ibid}. {\bf 72}, 6692 (1980).

\bibitem{XS:96}  K. Q. Xia, X.-Q. An, and W. G. Shen, {J. Chem. Phys.} {\bf %
105}, 6018 (1996).

\bibitem{XW:92}  K. Q. Xia, C. Frank, and B. Widom, {J. Chem. Phys.} {\bf 97}%
, 1446 (1992).

\bibitem{KD:00}  K. Kamide and T. Dobashi, {\it Physical Chemistry of
Polymer Solutions}, (Elsevier, Amsterdam, 2000).

\bibitem{PW:98}  A. Z. Panagiotoupoulos, V. Wong and M. A. Floriano, {%
Macromolecules} {\bf 31}, 912 (1998).
\end{references}
\end{document}